\documentclass{aa} 

\usepackage{siunitx}
\usepackage{graphicx}
\usepackage{txfonts}
\usepackage{todonotes}
\usepackage{booktabs}
\usepackage{hyperref}
\usepackage{listings}
\usepackage{natbib}
\usepackage{gensymb}

\begin{document}
\title{Closing the gap to convergence of gravitoturbulence in local simulations}
\titlerunning{}

\subtitle{}

\author{J.~Klee~\inst{1}\fnmsep\thanks{jklee@astrophysik.uni-kiel.de}
        \and
        T.~F.~Illenseer\inst{1}
        \and
        M.~Jung~\inst{2}
        \and
        W.~J.~Duschl\inst{1,3}
        }

\institute{Institut für Theoretische Physik und Astrophysik, Christian-Albrechts-Universität zu Kiel,
           Leibnizstr. 15, 24118 Kiel, Germany
           \and
           Hamburger Sternwarte, Universität Hamburg,
           Gojenbergsweg 112, 21029 Hamburg, Germany
           \and
           Steward Observatory, The University of Arizona,
           933 N. Cherry Ave., Tucson, AZ 85721, USA \\
           }

\date{\today, accepted for publication in A\&A}

\abstract
{}
{Our goal is to find a converged cooling limit for fragmentation in self-gravitating disks. This is
  especially interesting for the formation of planets, brown dwarfs or stars and the growth of black holes.
 While investigating the limit, we want to give a clear criterion for the state of convergence.}
{We run two-dimensional shearingsheet simulations with the hydrodynamic package \texttt{Fosite}
 at high resolutions. Thereby resolution and limiters are altered. Subsequently, we investigate the spectra of
 important physical quantities at the length scales where fragmentation occurs. In order
 to avoid prompt fragmentation at high resolutions we start these simulations with a fully
 developed gravitoturbulent state obtained at a lower resolution.
 }
 {We show nearly converged results for fragmentation with a critical cooling timescale
  $t_{\mathrm{crit}} \sim 10\,\Omega^{-1}$. We can backtrace this claim by investigating the spectra of relevant physical
  variables at length scales around and below the pressure scale height. We argue that
  well behaved results cannot be expected if counteracting quantities are varying too much on
  these critical length scales, either by change of resolution or numerical method. A comparison
  of fragmentation behaviour with the related spectra reveals that simulations behave similar, if
  the spectra are converged to the length scales where self-gravity leads to instabilities.
  Observable deviations in the results obtained with different numerical setup are confined
  to scales below these critical length scales.
 }
{}


\maketitle

\section{Introduction}\label{sec:intro}
Self-gravitating disks play an import role in the field of young protoplanetary disks and AGN-disks.
A disk becomes gravitationally unstable if the Toomre parameter
\begin{equation}
  Q = \frac{\kappa c_{\mathrm{s}}}{\pi G \Sigma},
\end{equation}
fulfils $Q < 1$ \citep{toomre1964gravitational}. Thereby, $\kappa$ is the epicyclic frequency, which
becomes the angular velocity $\Omega$ in case of a Keplerian disk, $c_{\mathrm{s}}$ is the sound speed,
$G$ the gravitational constant and $\Sigma$ the surface density. The Toomre criterion is viable in a
local approximation with axisymmetric potential.

Local gravitation act on a certain length-scale interval, where disturbances are allowed to grow
\citep{toomre1964gravitational,lin1987viscosity}. It is determined by the critical length scale as an
upper limit $L_{\mathrm{crit}}=G \Sigma/ \Omega^2$ and the Jeans length scale
$L_{\mathrm{J}} = c_{\mathrm{s}}^2/ G \Sigma$ as a lower limit. Thus perturbances within
\begin{equation}\label{eq:lengthregion}
  L_{\mathrm{J}} \le L \le L_{\mathrm{crit}}
\end{equation}
are amplified and could eventually lead to fragmentation. \citet{nelson2006numerical} imposed numerical
consequences for this and stated that the Toomre-wavelength
$\lambda_{\mathrm{T}} = 2L_{\mathrm{J}} \sim H$ needs to be resolved by a simulation,
where $H=c_{\mathrm{s}} / \Omega$ is the Keplerian scale height.

\citet{gammie2001nonlinear} showed in his paper numerically that a slow-cooling disk
settles into the gravitoturbulent state \citep{paczynski1978model}, whereas a gravitationally
unstable disk that cools fast leads to fragmentation. In the former case turbulence
leads to an enhanced transport of angular momentum
\citep{lin1987viscosity,balbus1999dynamical,duschl2000note,duschl2006gravitational}.
In the latter the gaseous disk may at least partly disintegrate if giant planets, brown
dwarfs or, in case of AGN disks, even massive stars form within the disk
\citep{rice2015disc,baehr2017fragmentation,boss1998evolution,levin2003stellar}.
The criterion for fragmentation that \cite{gammie2001nonlinear} ascertains
is that the cooling fulfils $\beta = t_{\mathrm{c}}\Omega < 3$, where $\beta$ is a dimensionless quantity
describing the cooling timescale $t_{\mathrm{c}}$ in terms of orbital timescale $\Omega^{-1}$.
The value for $\beta$ can also be related to an effective viscosity parameter
$\alpha = (\gamma \left( \gamma - 1\right) q^2 \beta )^{-1}$
\citep{gammie2001nonlinear,shakura1973black}. Here, $\gamma$ is the ratio
of specific heat capacities and $q = -\frac{\partial \ln{\Omega}}{\partial \ln{r}}$ is
the power-law exponent of the rotation law which becomes $3/2$ for
Keplerian motion.

However, simulations show also fragmentation for larger $\beta$, while going to higher resolutions
\citep{meru2011non,paardekooper2011numerical}. This non-convergence also takes place in the local-approximation
\citep{paardekooper2012numerical,baehr2015role,klee2017impact}. Even by using an altered cooling at small scales,
\citet{baehr2015role} suffer from fragmentation in their
high-resolution runs. Recently, \citet{deng2017convergence} show that codes that rely on smoothed particle
hydrodynamics that use artificial viscosity also have problems with artificial fragmentation. This can be circumvented
by using meshless methods. But also grid based codes with shock capturing mechanisms can get into
troubles, where limiting functions can have a strong impact on the outcome of the simulations \cite{klee2017impact}.

\citep{young2015dependence} also observed fragmentation above the critical $\beta_{\mathrm{crit}}$ in their simulations.
They proposed that a quasi-static collapse can explain this behaviour.
\cite{paardekooper2012numerical} shows that there is also a stochastic component when running
over long timescales.

More recently, also three dimensional shearingbox simulations were carried out in the local approximation
\citep{shi2014gravito}. \citet{baehr2017fragmentation} found fragmentation
at around $\beta \lesssim 3$ reaching high resolutions with very small box-sizes.
\citet{riols2017gravitoturbulence} show that at sufficiently high resolutions small scale
turbulences set in, off the midplane. They furthermore point out that the
energy spectrum of gravitoturbulence can be approximated by the $k^{-5/3}$ cascade \citep{kolmogorov1941local}.
However, they raise the problem already revealed by \citep{gammie2001nonlinear} that in order to have a well-behaved gravitoturbulent state
the box size should be not too small. They conclude a box size of at least $\gtrsim 40\,H = 40 \pi L_{\mathrm{crit}}$.
Otherwise bursts that might be connected to transient fragmentation set in. \citet{booth2018characterizing}
propose a minimal size of the shearingbox $\gtrsim 64\,H$ in horizontal direction. Otherwise long term trends show up.

In this paper we investigate fragmentation at very high horizontal resolutions while preserving a well-behaved
gravitoturbulent state in a sufficiently large shearing sheet.
We first describe our methods in section~\ref{sec:methods}, explaining the underlying
equations and the setup. In section~\ref{sec:results}
we present the numerical results obtained for highest resolutions and closed to converged
spectra for various quantities. Finally, we discuss and conclude on our results in section \ref{sec:discussion}.

\section{Methods}\label{sec:methods}
We use the hydrodynamic code \texttt{Fosite} by \cite{illenseer2009two}, a generalized version of the finite volume
method of \citet{kurganov2000new}. \texttt{Fosite} can operate on arbitrary curvilinear-orthogonal
grids and is fully parallelized and vectorized. It uses modern Fortran 2008 language features and
object-oriented design patterns. Fosite has recently been ported and optimized for the new NEC SX-Aurora
Tsubasa vector computer which gave us the ability to carry out simulations with unprecedented high
resolution in two dimensions\footnote{The code is freely available under \url{https://github.com/tillense/fosite}.}.

\subsection{Hydrodynamic Equations}\label{sec:hydroeq}
We solve the two-dimensional hydrodynamical equations in the local approximation \citep{goldreich1965ii,hawley1995local}.
Together with the Poisson equation for a razor thin disk \citep{gammie2001nonlinear} the system of equations is
\begin{subequations}
\label{eq:hydro}
\begin{align}
    \frac{\partial \Sigma}{\partial t} + \nabla \cdot \left( \Sigma \mathbf{v} \right) - q\Omega y \frac{\partial
    \Sigma}{\partial x} = &\, 0 \label{eq:hydro_a},\\
    \frac{\partial \Sigma \mathbf{v}}{\partial t} + \nabla \cdot \left(\Sigma
    \mathbf{v}\otimes \mathbf{v} + p \mathbb{I} \right) - q \Omega y \frac{\partial \Sigma \mathbf{v}}{\partial x}= &\,
		\Omega \Sigma \begin{pmatrix} (2-q) v_y \\ -2 v_x \end{pmatrix} - \Sigma \nabla \Phi_{\mathrm{sg}} \label{eq:hydro_b},\\
    \frac{\partial E}{\partial t} + \nabla \cdot \left( \left(E + p \right) \mathbf{v} \right) - q\Omega y
    \frac{\partial E}{\partial x}= & \, -q \Omega \Sigma v_x v_y \nonumber \\
    & - \Sigma \nabla \Phi_{\mathrm{sg}} \cdot \mathbf{v} - \Sigma Q_{\mathrm{cool}}, \label{eq:hydro_c} \\
		\Delta \Phi_{\mathrm{sg}} = & \, 4\pi G \Sigma \delta(z). \label{eq:poisson}
\end{align}
\end{subequations}
Here, $p$ is the integrated pressure and $\mathbf{v} = v_x \mathbf{\hat{e}}_x + v_y \mathbf{\hat{e}}_y$
is the residual part of the two dimensional velocity, by splitting of a background velocity $q\Omega y \mathbf{\hat{e}}_x$.
The according total energy is $E = \frac{1}{2} m |\mathbf{v}|^2 + p/(\gamma - 1)$. The additional
advection step, which is the last term on the left hand side of eqs.~\ref{eq:hydro_a} - \ref{eq:hydro_c}
is calculated separately in order to allow for very large time steps \citep{masset2000fargo,jung2018multi}. The gravitational potential
of the shearingsheet $\Phi_{\mathrm{sg}}$ describes the self-gravitation.
Thereby, $\delta(z)$ is the delta-distribution.
The according Poisson equation~\ref{eq:poisson} is solved in Fourier space \citep{gammie2001nonlinear,klee2017impact},
imposing periodic boundary conditions. This is achieved by shifting the whole density field
to the next periodic point before applying the transformation \citep{gammie2001nonlinear,klee2017impact}.
Finally, $Q_{\mathrm{cool}} = \frac{p \Omega}{(\gamma -1) \beta}$ is the well-known cooling
parametrization \citep{gammie2001nonlinear}.

\subsection{Numerical Setup}
In order to have a well-tuned setup we rely on the results in \citet{klee2017impact}.
There different numerical shock-capturing mechanisms (limiting functions) are tested
against each other. We thus prefer van-Leer limiting \citep{vanleer1974towards}, which
showed to be advantageous in contrast to other limiting functions, regarding
artificial fragmentation. However, most simulations were again carried out using the
Superbee limiter \citep{roe1982algorithms} for comparison. Additionally, we use Dormand-Prince
time stepping \citep{dormand1980family}, which showed the best
results in the epicyclic motion test \citep{klee2017impact}.

We rotate the computational domain by $90\degree$ in contrast to \cite{klee2017impact}. This
is advantageous in order to have the first dimension
completely accessible for the parallel Fourier solver, while preserving vectorization. We
thus decomposed the grid in pencils along the $y$-direction.

\subsubsection{Simulation setup}
With $L_x = L_y = 320\, L_{\mathrm{crit}}$ we use the same grid size as in \citet{gammie2001nonlinear}. This
corresponds to $\sim 100\,H$. It ensures that we are not getting into trouble with long-term periodics as
stated in \citet{booth2018characterizing}. Another reason for staying with a relatively large extent
of the box compared to other three-dimensional simulations is already given by \citet[Fig.~9]{gammie2001nonlinear}.
There, the contribution of the gravitational component of the stress is taken into consideration for twice of our
size $L = 640\, L_{\mathrm{crit}}$. Taking the length of $L = 320\, L_{\mathrm{crit}}$ into account,
still includes $\sim 90 \%$ of the gravitational component. Further reduction of box size removes a
significant amount of gravitational stress from the simulations which may have a major impact on the
results. This is supported by the observations of \citet{riols2017gravitoturbulence} and \citet{booth2018characterizing}
who found that whereas some parameters, like Toomre's $Q$ converge
quite fast for increasing box sizes, kinetic energy and gravitational stress are not doing so.

We use $\gamma=2.0$ in order to conform to previous results
\citet{gammie2001nonlinear,paardekooper2012numerical,klee2017impact}. It should be kept in
mind, however,that smaller $\gamma$ generally increase the probability for fragmentation
\citep{rice2016evolution}.

\citet{booth2018characterizing} investigated in more detail the problem of prompt fragmentation, where
fragments form during initialization. In
\citet{klee2017impact} we already used the approach of slowly moving down the cooling in order
to prevent this kind of fragmentation also in local simulations.
However, going to higher resolutions made it increasingly harder for us to prevent the artificial
fragmentation in this manner. Additionally, reaching the gravitoturbulent state needs more
and more time, for higher resolutions and slower (initial) cooling.

In this work, we take another approach. Instead of fine-tuning the initial cooling, we reuse
the result of a previous fully developed low resolution run with slow cooling using
spline-interpolation to generate high resolution data\footnote{RegularGridInterpolator
of the SciPy package \citep{scipy2019}.}.
Thereby we already initiate the high resolution simulation with a gravitoturbulent state.
This approach also saves a significant amount of computational time skipping the
initial phase in which the simulation settles into the gravitoturbulent state.


\section{Simulations and results}\label{sec:results}
\begin{table}[htbp]
  \caption{Overview of simulation runs}
  \label{tab:simulations}
  \centering
  \begin{tabular}{ccccccc}
    \hline\hline
      Name        & Lim.& Res. 	 & $\beta$  & $T_{\mathrm{sim}}\; \Omega$ & Frag. \\
    \hline
      VL4096B5    & VL  & $4096$ & $5$      & $50$                        & Yes   \\
      VL4096B7-5  & VL  & $4096$ & $7.5$    & $175$                       & Yes   \\
      VL4096B10   & VL  & $4096$ & $10.0$   & $200$                       & Yes   \\
      VL4096B15   & VL  & $4096$ & $15.0$   & $200$                       & No    \\
      VL4096B20   & VL  & $4096$ & $20.0$   & $200$                       & No    \\
      VL8192B5    & VL  & $8192$ & $5.0$    & $34$                        & Yes   \\
      VL8192B10   & VL  & $8192$ & $10.0$   & $120$                       & Yes   \\
      VL8192B15   & VL  & $8192$ & $15.0$   & $120$                       & No    \\
      VL8192B20   & VL  & $8192$ & $20.0$   & $100$                       & No    \\
      SB4096B5    & SB  & $4096$ & $5$      & $50$                        & Yes   \\
      SB4096B7-5  & SB  & $4096$ & $7.5$    & $100$                       & Yes   \\
      SB4096B10   & SB  & $4096$ & $10.0$   & $250$                       & Yes   \\
      SB4096B15   & SB  & $4096$ & $15.0$   & $200$                       & No    \\
      SB4096B20   & SB  & $4096$ & $20.0$   & $200$                       & No    \\
      SB8192B5    & SB  & $8192$ & $5.0$    & $100$                       & Yes   \\
      SB8192B7-5  & SB  & $8192$ & $7.5$    & $50$                        & Yes   \\
      SB8192B10   & SB  & $8192$ & $10.0$   & $100$                       & No    \\
    \hline
  \end{tabular}
  \tablefoot{VL - van-Leer, SB - Superbee, $T_{\mathrm{sim}}$ - simulation time}
\end{table}
Table~\ref{tab:simulations} shows an overview of the simulations that were done. We ran
simulations at twice and four times the highest resolution in \citet{klee2017impact},
which corresponds to a resolution of $4096$ and $8192$, respectively.

As initial conditions for the runs with resolution $4096$, we used a gravitoturbulent
run with a resolution of $2048$, van-Leer limiting and $\beta=10$. Since our run at $\beta=10$
fragmented at a resolution of $4096$ we used the run with $\beta=15$ as a starting point for
the runs with a resolution of $8192$. At a resolution of $4096$ the normal runtime is
$200\, \Omega^{-1}$ and at $8192$ it is
$100\, \Omega^{-1}$. If many clear fragments formed we stopped the run earlier. For
small density enhancements at the end of the run we optionally
extended the run to investigate its outcome.
\begin{figure}[htbp]
    \centering
    \includegraphics[width=\hsize]{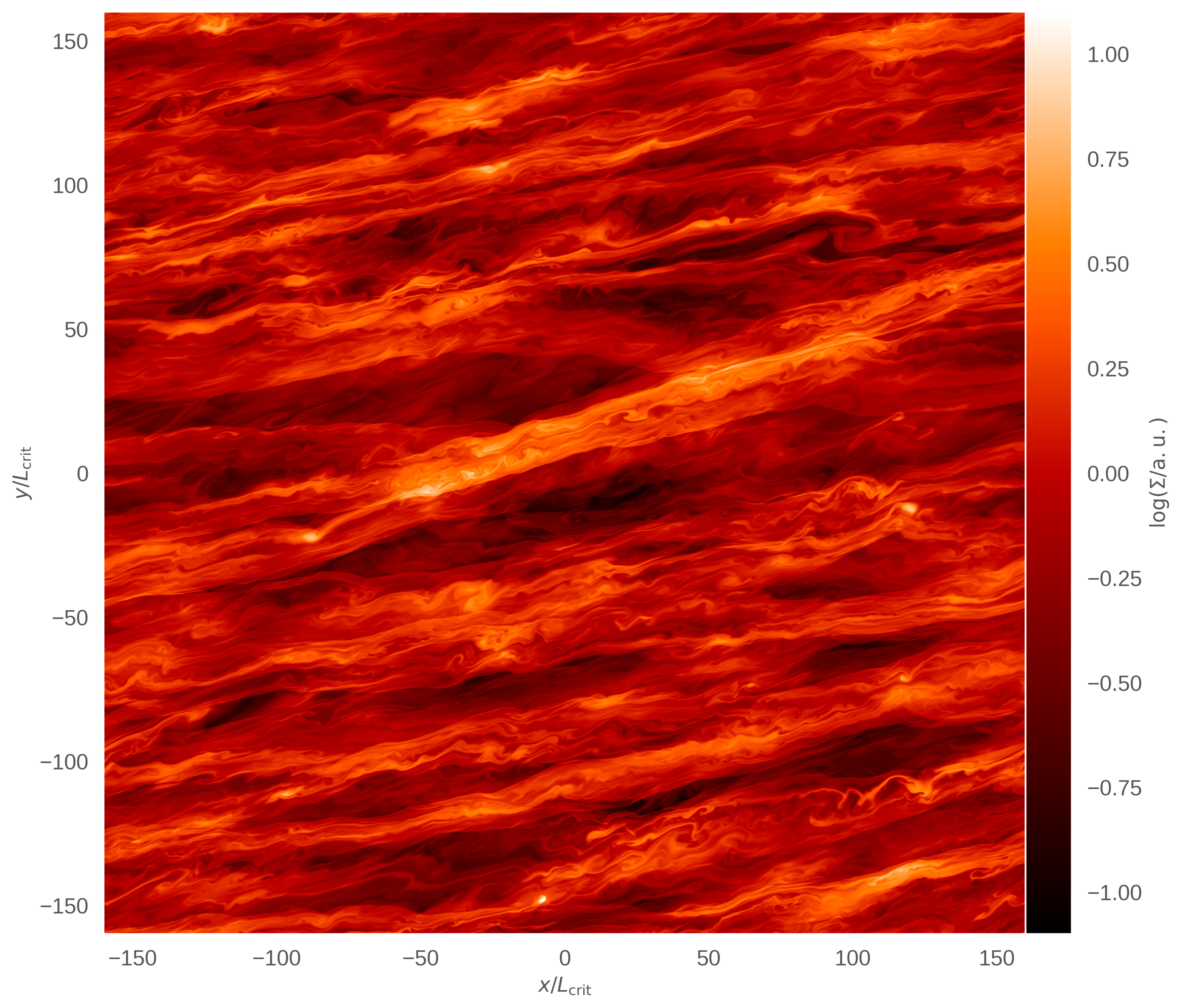}
    \caption{Snapshot of the surface density of the gravitoturbulent state
		at a resolution of $8192$ for $\beta=10$ and van-Leer limiting. This simulation
		shows later fragments with $20-30$ times the background density.}
    \label{fig:gravitoturbulence_VL}
\end{figure}
\begin{figure}[htbp]
    \centering
    \includegraphics[width=\hsize]{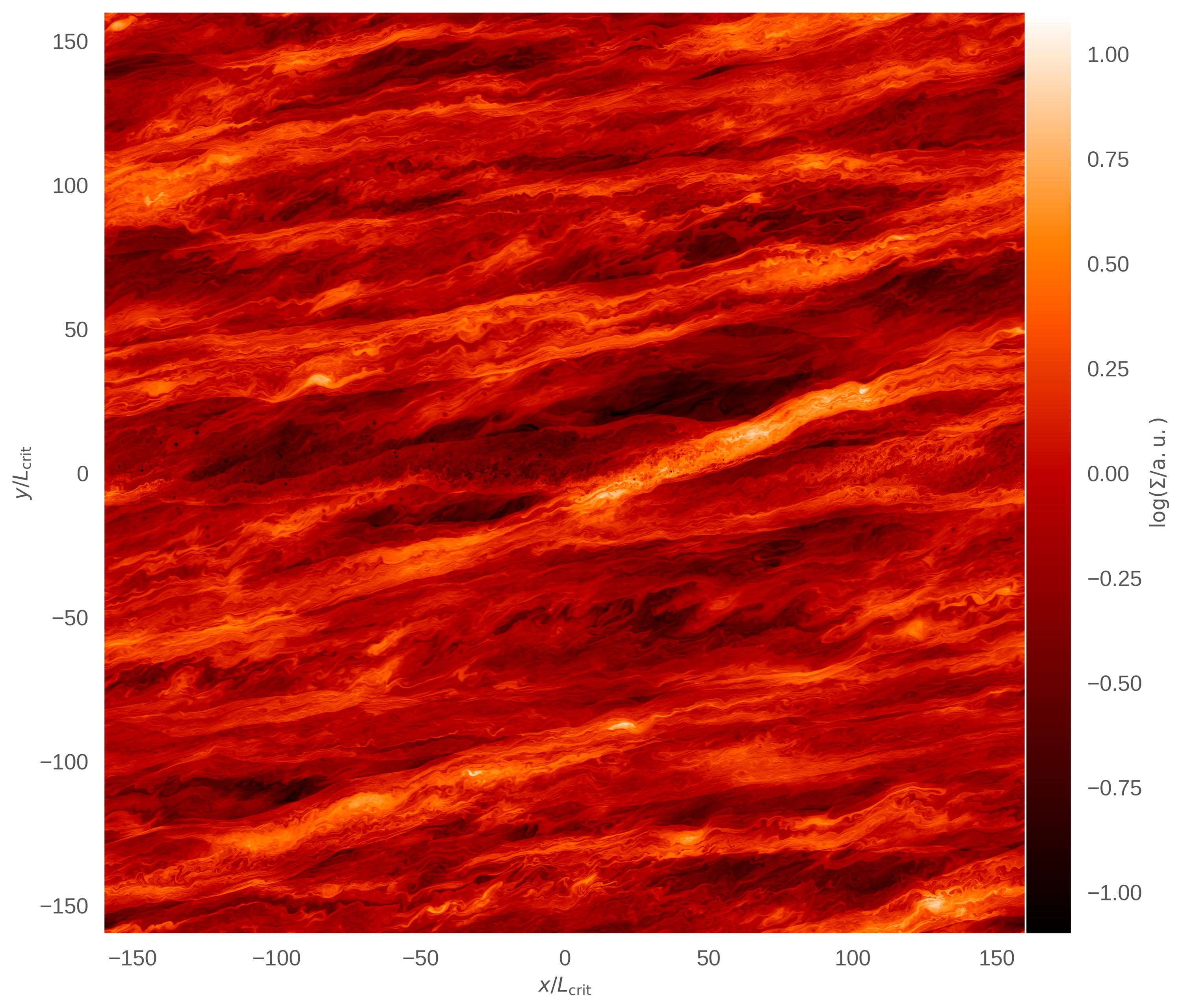}
    \caption{Snapshot of the surface density of the gravitoturbulent state at a
		resolution of $8192$ for $\beta=10$ and Superbee limiting.}
    \label{fig:gravitoturbulence_SB}
\end{figure}
\begin{figure}[htbp]
    \centering
    \includegraphics[width=\hsize]{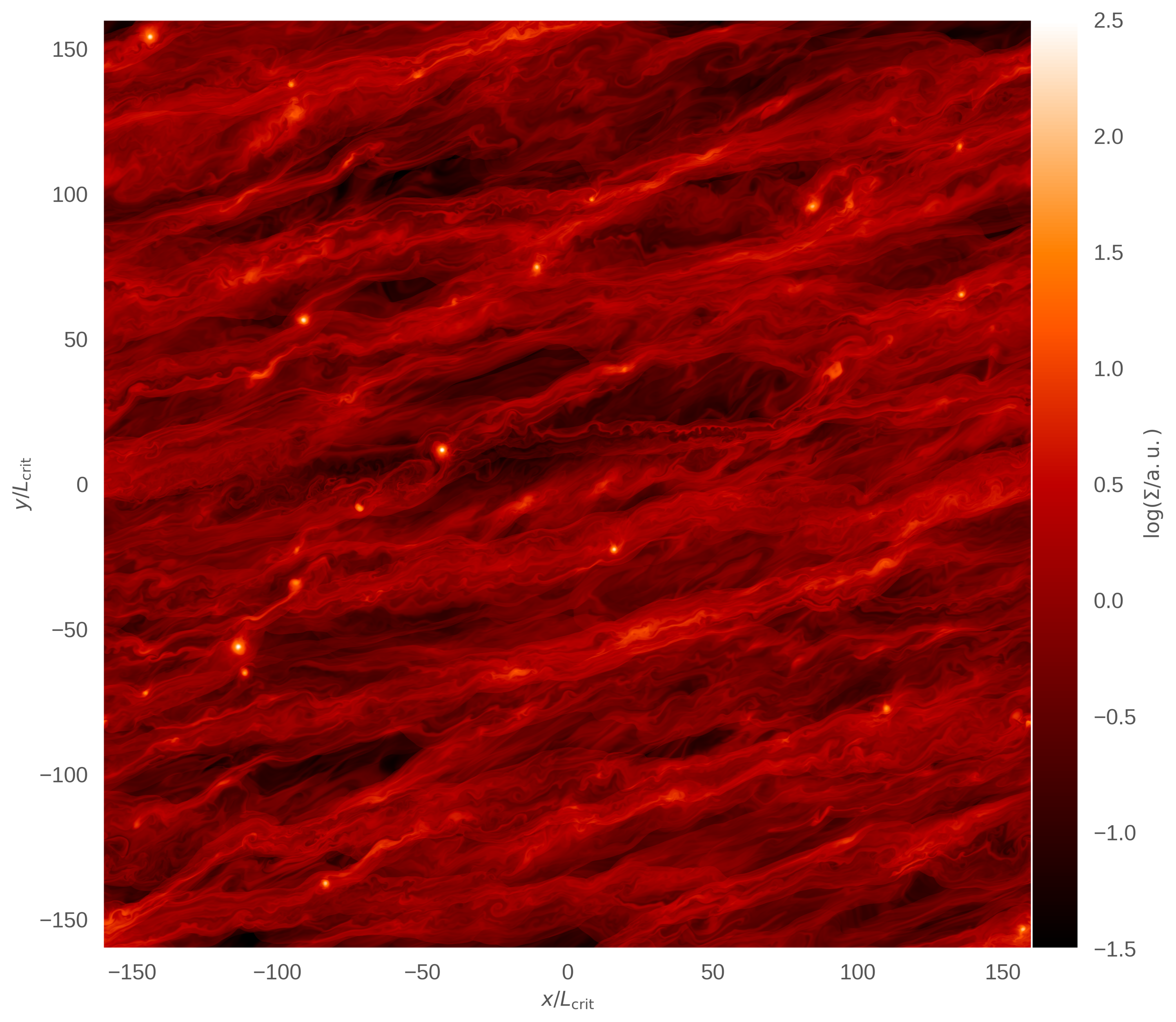}
    \caption{Snapshot of the surface density in a fully fragmented run at a
		resolution of $8192$ for $\beta=5$ and van-Leer limiting.}
    \label{fig:fragmentation}
\end{figure}

In case of simulations carried out with the Superbee limiter, a close inspection of the
region within $y=\pm 20\,L_{\mathrm{crit}}$ reveals small almost circular features with unusual
density depletion which were never observed in simulations with van-Leer limiter. It is not clear
to us what determines the spatial occurrence and extent as well as the life time of these structures.
As a matter of fact they seem to be continuously generated only around the midplane with scales of the
order of at most a few $L_{\mathrm{crit}}$ and are therefore only visible in the simulations at resolutions
$4096$ and $8192$. If they encounter strong shocks they are often completely destroyed but some of them survive
for several orbital time scales.

We think that these structures are numerical artifacts again caused by oversteepening \citep{klee2017impact}.
Since the Superbee limiter tends to overestimate spatial gradients it amplifies low-pressure areas. The results
shown for the isentropic vortex test \citet{yee1999low} in App.~\ref{app:vortex_test} support this hypothesis.
They clearly illustrate that the Superbee limiter enhances the central pressure trough while the van-Leer limiter depletes it.

Fig.~\ref{fig:gravitoturbulence_VL}, \ref{fig:gravitoturbulence_SB} and \ref{fig:fragmentation} show
snapshots of the high-resolution results obtained with different limiters and cooling parameters. In case
of the gravitoturbulent simulations we also provide movies for better illustration of what is going on in
these simulations and to make the differences between the two limiters clearer.

\subsection{Fragmentation}
\begin{figure}[htbp]
  \centering
  \includegraphics[width=\hsize]{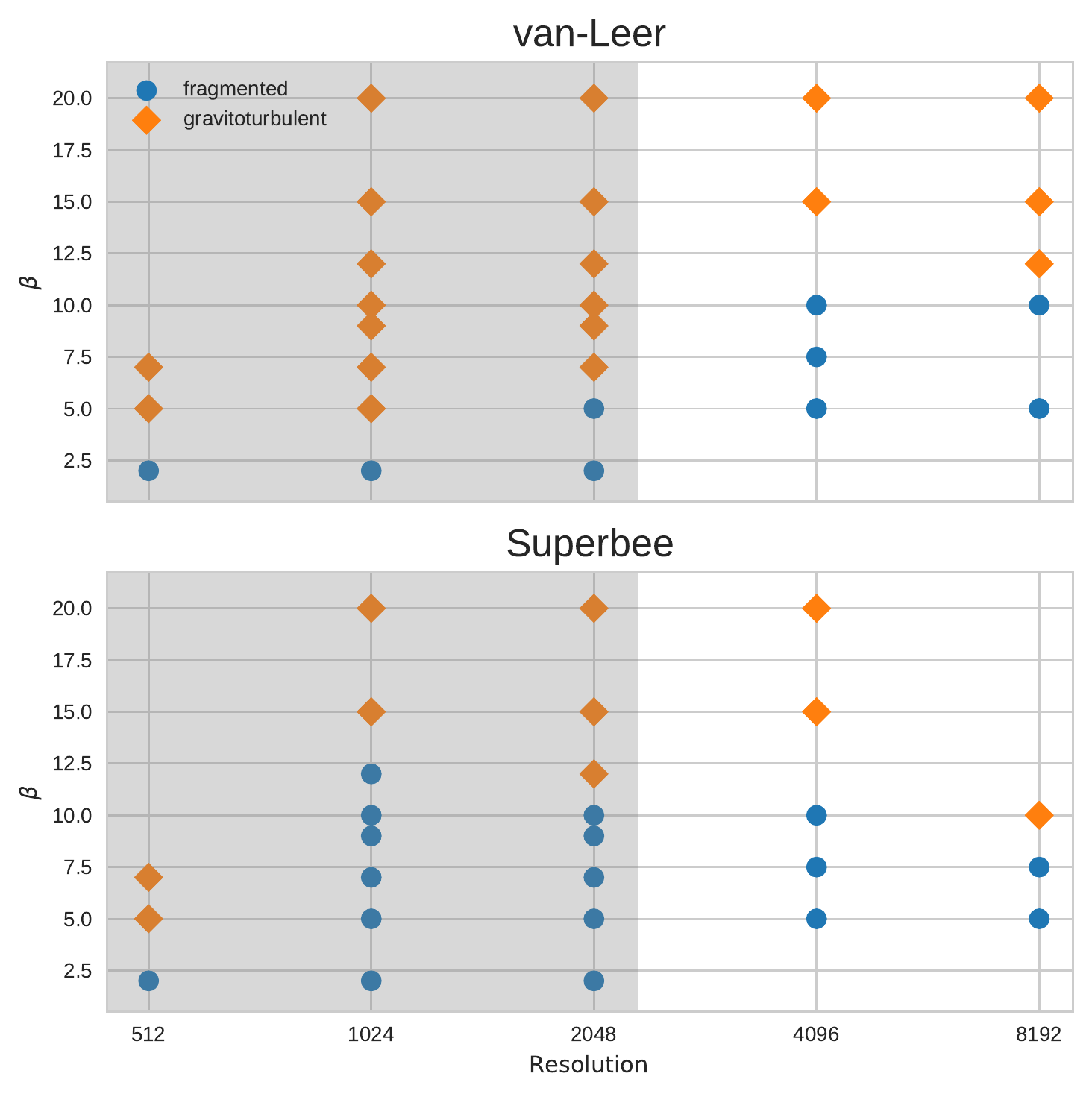}
  \caption{Fragmentation plot for different resolutions $N$ and cooling timescales $\beta$. The
    gray regions show results already published in \citet{klee2017impact}. The simulations seem
    to converge at the highest resolutions to a value of $\beta_{\mathrm{crit}}
    \approx 10$ for both numerical schemes.}
  \label{fig:fragplot}
\end{figure}
Fig.~\ref{fig:fragplot} summarizes which simulations showed fragmentation. Besides the
new runs we also added the results for lower resolutions from \citet{klee2017impact}. For
higher resolutions the question whether or not fragmentation occurs is less affected by the
limiter. This is not what
we would expect in view of our argumentation in \citet{klee2017impact}, where Superbee suffers
from numerical errors that force fragmentation. Looking at even higher resolutions, we see that
the Superbee limiter does not fragment for $\beta=10$. Thus in contrast to our previous
findings it seems that the Superbee limiter yields similar results with respect to the
fragmenting behaviour and the threshold for fragmentation stabilizes around $\beta \sim 10$.

For $\beta=10$ and van-Leer limiting at
$N=4096$ we have a single fragment that forms between $50\,\Omega^{-1}$ and $75\,\Omega^{-1}$ and
is disrupted again before the end of the run. On the other side we had a fragment for
the same setup with Superbee at the regular end of the simulation. We then extended the
simulation another $50\,\Omega^{-1}$ in order to ensure that it is stable for a longer
period. A similar behaviour was found in the run with $N=8192$ for van-Leer limiting and
$\beta=10$. In this case a fragment forms around $\sim 80\,\Omega^{-1}$ which has $20-30$ times
the background density and is stable for at least $\sim 40\, \Omega^{-1}$. Thus only
single or few fragments form in the range $5 < \beta \lesssim 10$ whereas below $\beta=5$ the
whole field fragments immediately (cf. Fig.~\ref{fig:fragmentation}). In addition, if
$\beta\sim 10$ these single fragments do sometimes not survive for longer periods.
Therefore the conclusion whether or not the simulation is fragmented seems less clear in these
cases since it depends on the formation and stability of a single fragment.

\subsection{Convergence of spectra at important length scales}
\begin{figure}[htbp]
    \centering
    \includegraphics[width=\hsize]{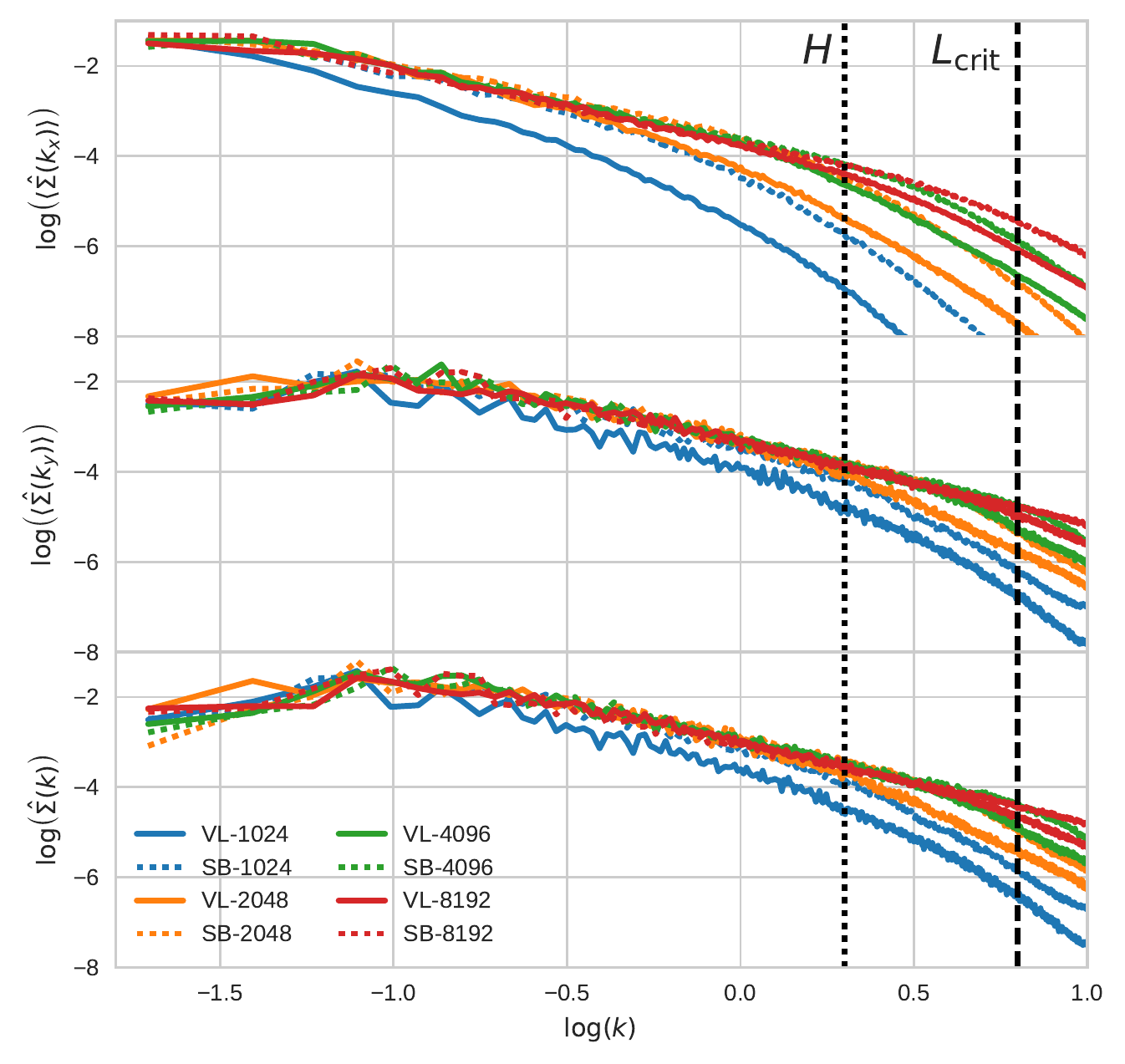}
    \caption{Components of the density spectrum for different resolutions and limiters.
		The density spectrum is well converged at highest resolutions to a length scale of $H$.
    The component in $x$-direction is only close to convergence down to $L_{\mathrm{crit}}$.}
    \label{fig:density_spectrum}
\end{figure}
\begin{figure}[htbp]
    \centering
    \includegraphics[width=\hsize]{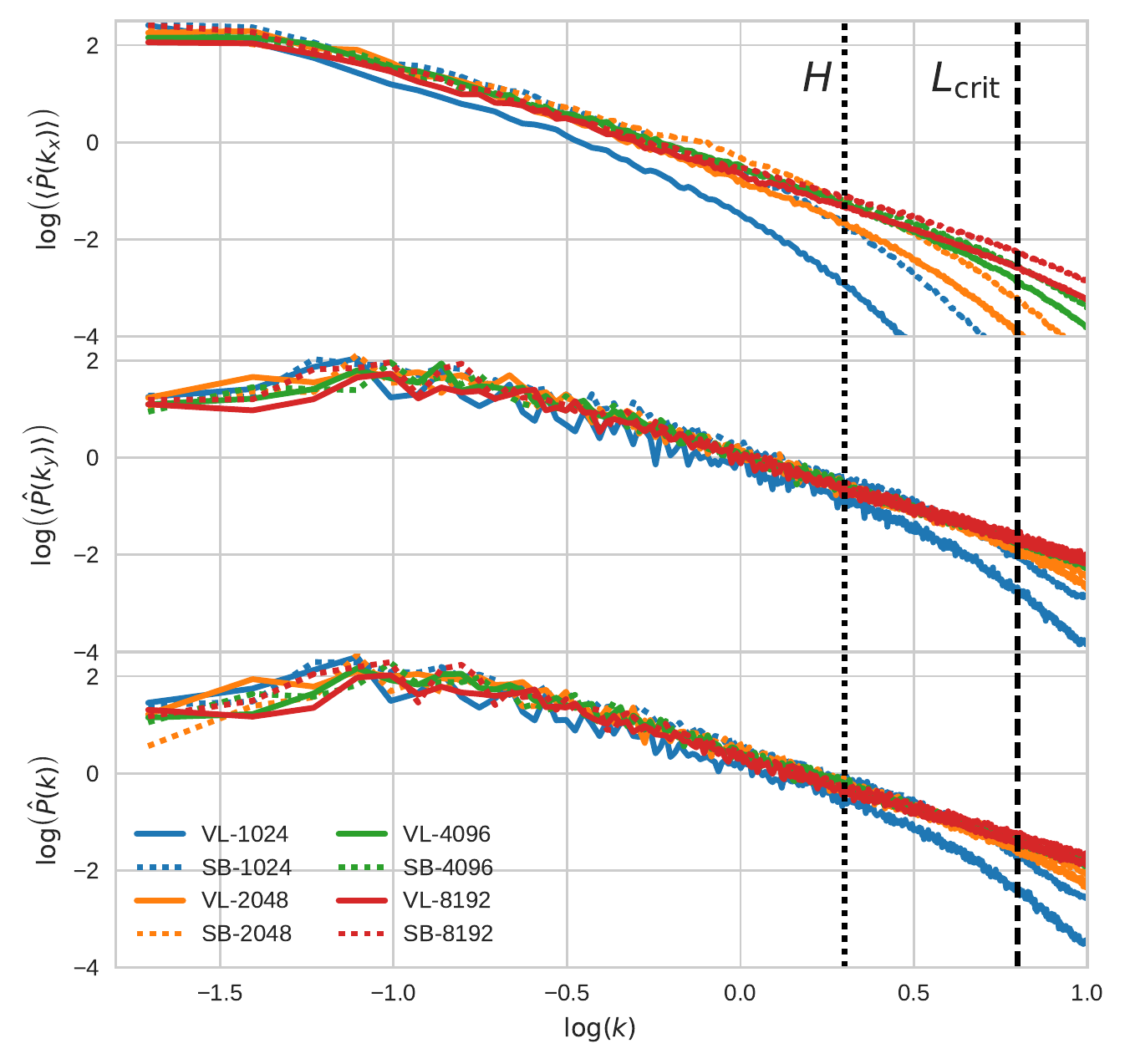}
    \caption{Components of the pressure spectrum for different resolutions and limiters.
		The pressure spectrum is well converged at highest resolutions even to a length scale
    of $L_{\mathrm{crit}}$.}
    \label{fig:pressure_spectrum}
\end{figure}
In Fig.~\ref{fig:density_spectrum} and Fig.~\ref{fig:pressure_spectrum} we show density
and pressure spectra, respectively. The spectra are adopted for a cooling parameter of
$\beta=10$ at a state where fragmentation did not occur, yet (if present). Thereby, their components
in $k_x$, $k_y$ and $k$ are displayed. We see most pronounced in the surface density that
the results obtained with low resolutions deviate significantly from those at higher
resolutions even on length scales of the order of the scale height and above.
The plots show that runs at the same resolution but with different limiters can
lead to a deviation in the spectra that is partially larger than an order of magnitude. The
difference between resolutions is here of several magnitudes.
At higher resolutions we see that the spectral quantities are very
similar and follow the cascade down to the scale height $H$. For many quantities this
is even true for the smaller length scale $L_{\mathrm{crit}}$, especially at the resolution
of $8192$. The $x$-components, which correspond in our case to the direction which is
more parallel to the shock fronts, lacks a bit of convergence at least at the length-scale
$L_{\mathrm{crit}}$.

\begin{figure}[htbp]
    \centering
    \includegraphics[width=\hsize]{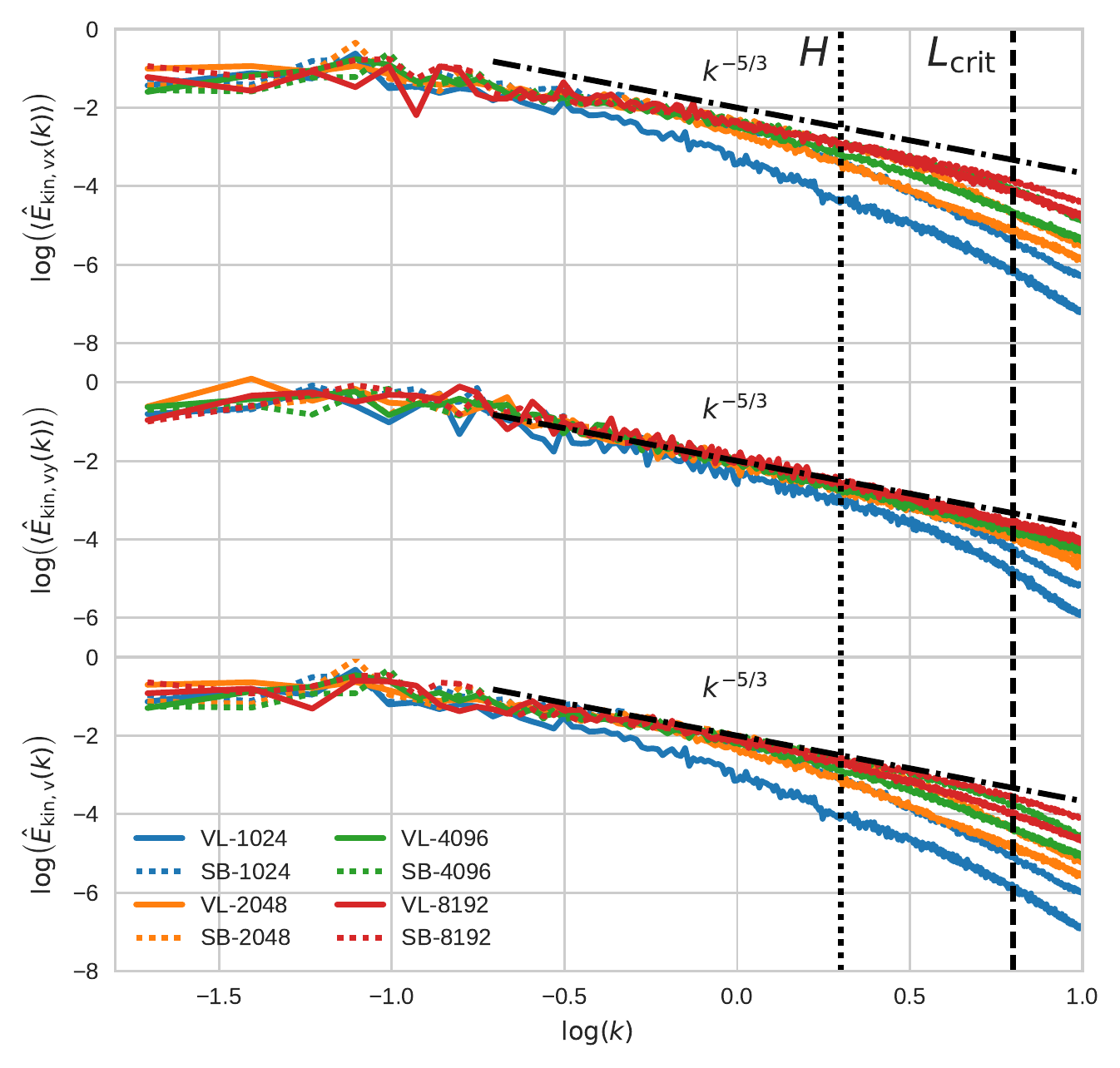}
    \caption{Spectrum of kinetic energy for different resolutions and limiters. The velocities
             are summed in Fourier space along the radius $k$. The spectrum fulfils the
             $k^{-5/3}$ well at highest resolution down to $L_{\mathrm{crit}}$.}
    \label{fig:energy_spectrum}
\end{figure}
In Fig.~\ref{fig:energy_spectrum} we show spectra of the kinetic energy. In contrast to
Fig.~\ref{fig:density_spectrum} and \ref{fig:pressure_spectrum} we do not show the components
along $k_x$ or $k_y$ but the kinetic energy associated with $v_x$, $v_y$ and $|\mathbf{v}|$,
always summed up along the radius in Fourier space $k$. We see in
Fig.~\ref{fig:energy_spectrum} the $\propto k^{-5/3}$-law fairly
good, which is expected in two dimensions at large scales \citep{kraichnan1967inertial}.

All figures mentioned above show well-converged spectra at the scale
height $H$ at high resolutions. Additionally, they even have few differences at
$L_{\mathrm{crit}}$. Having a look again at Fig.~\ref{fig:fragplot} we can state that the
simulations that show differences with respect to fragmenting behaviour also
show differences for the physical quantities at the length scales where
fragmentation occurs. When this gap is closed, we also see smaller differences in the
fragmenting behaviour. We thus have a clear criterion for convergence, which does
not only rely on the observation of forming clumps, but on the turbulent cascade
down to the relevant length scales.

\section{Discussion \& Conclusion}\label{sec:discussion}
We show results of self-gravitating shearing sheet simulations conducted in two dimensions
at the highest resolution so far and investigate the fragmentation behaviour.
In order to overcome prompt fragmentation we interpolate from an already evolved gravitoturbulent
state with lower resolutions and use this as initial conditions. However, even then
we get a relatively large critical value of $\beta_{\mathrm{crit}} \lesssim 10$ as
a fragmentation criterion. We regard this result
as robust and independent of resolution as long as it exceeds $4096$ at the least
($40$ cell per scale-height $H$). Still, there seems to be some uncertainty around the
aforementioned $\beta_{\mathrm{crit}}$ swiftly declining for higher values.
This holds also for different numerical schemes within the code, where we know that
one of the limiting functions is prone to numerical errors. However, even with
these errors the critical value $\beta_{\mathrm{crit}}$ is not affected too much.

That is why we think that we are close to convergence. As a definition for that state
we do not only have a look at the fragmentation behaviour with ever higher resolutions.
Instead we investigate physical quantities of interest at the length scales where fragmentation
occurs. When resolutions are high enough, the small scale structures show very clear cascades
down to the length-scales of interest in eq.~\ref{eq:lengthregion}. Although even
smaller structures can vary fundamentally between the numerical schemes, we get
similar results for fragmentation. We conclude that the differences in the
smallest structures have (nearly) no influence on the fragmentation process.
This is the case, because the amplified modes by self-gravity only affect
the length scales in eq.~\ref{eq:lengthregion}.

Recent results preferred a clearer criterion at around $\beta_{\mathrm{crit}}=3$
\citep{baehr2017fragmentation,booth2018characterizing,deng2017convergence}. They all have the
advantage that they were run in three dimensions. However, in order to reach the necessary
high horizontal resolutions \citet{baehr2017fragmentation} and \citet{booth2018characterizing}
reduce the spatial dimensions ($12\,H$ and max. $32\,H$) for the runs in which they
investigate fragmentation. Indeed we see that the runs where \citet{booth2018characterizing}
resolve $H$ by $32$ grid cells show fragmentation for slightly larger $\beta_{\mathrm{crit}}\sim 4$.
Interestingly, this corresponds to our runs between resolutions of $2048$ and $4096$
(cf.~Fig.~\ref{fig:fragplot}). \citet{deng2017convergence} also show that they see no
fragmentation at higher resolutions with the meshless method code \texttt{Gizmo}
\citep{hopkins2014gizmo}. They state this is due to the lack of artificial viscosity in these
schemes. However, their code introduces slope limiting procedures, similar to those used in
our code Fosite which evidentially impact the outcome of fragmentation studies \citep{klee2017impact}.

It should also be taken into account, that two-dimensional turbulence can
behave differently when compared to three-dimensional turbulence. It has an
additional turbulent energy
cascade at small scales with $\propto k^{-3}$ \citep{kraichnan1967inertial,boffetta2012two,kolmogorov1941local}.
We see a fairly clear $k^{-5/3}$ cascade down to $L_{\mathrm{crit}}$
(cf.~Fig.~\ref{fig:energy_spectrum}). On scales below $L_{\mathrm{crit}}$, however,
the flow is no longer turbulent, because gravitational instabilities are not amplified
anymore. Nevertheless, it would be interesting to carry out a similar analysis of converged
spectra in three-dimensional simulations and verify these considerations.

The larger value of $\beta_{\mathrm{crit}}$ has some implications for fragmenting disks.
Relating the critical cooling parameter to the effective viscosity parameter $\alpha$
yields a smaller critital value for this number around
$\alpha_{\mathrm{crit}} \sim 0.02$ \cite{gammie2001nonlinear}.
This would indeed mean that $\alpha$-parametrizations with larger values might not
be well-suited in case of self-gravitating disks.

This reasonening becomes less clear if one takes additional heat sources into account.
For example, \citet{rice2011stability} show that an additional background irradiation cannot
suppress fragmentation, but leads to fragmentation at smaller $\beta_{\mathrm{crit}}$ with smaller
values for $\alpha_{\mathrm{crit}}$. The former direct relation between the two parameters is then
disconnected. Even when the cooling is fast, the reachable value for $\alpha_{\mathrm{crit}}$
becomes smaller.

In case of protoplanetary disks the shift to a larger value of a few for
$\beta_{\mathrm{crit}}$ has not necessarily a strong impact on the radial distance for
fragmentation in a global setup. Generally, for larger $\beta_{\mathrm{crit}}$
fragmentation can occur closer to the central object with a relation of
$\beta \propto r^{-9/2}$ \citep{clarke2009limits}. The cooling parameter thus
strongly depends on the radius \citep{paardekooper2012numerical}, which leads to
little effect in radial change. The distance of formation would than still be at
a radius of some ten astronomical units \citep{meru2012convergence}.
Regarding the lifetime of self-gravitating protoplanetary disks of
$\sim 10^5\,\mathrm{yr}$ \citep{laughlin1994nonaxisymmetric,haisch2001disk}
the results are still relevant. Taking into consideration a radius of
$r=100\,\mathrm{au}$ the disk would survive for $\sim 100$ orbits whereas
the simulations of this work were done for up to $200\,\Omega^{-1}$ which would
account for $\sim 32$ orbits.

Finally, we come to the conclusion that while going to ever higher resolution we inevitably
see again a shift to a larger $\beta_{\mathrm{crit}}\sim 10$ with some uncertainty around
that value. While for $\beta \lesssim 5$ the whole field fragments, we see only single or a few
fragments around $5 < \beta \lesssim 10$. Because of the close to converged spectra with different
numerical setups we do not expect this upper limit to change too much anymore.

\begin{acknowledgements}
We thank the anonymous referee for the very helpful comments and suggestions on a more recent
version of this publication.
\end{acknowledgements}

\bibliographystyle{aa}
\bibliography{references}

\appendix

\section{Vortex Test Cuts}\label{app:vortex_test}
\begin{figure}[htbp]
    \includegraphics[width=\hsize]{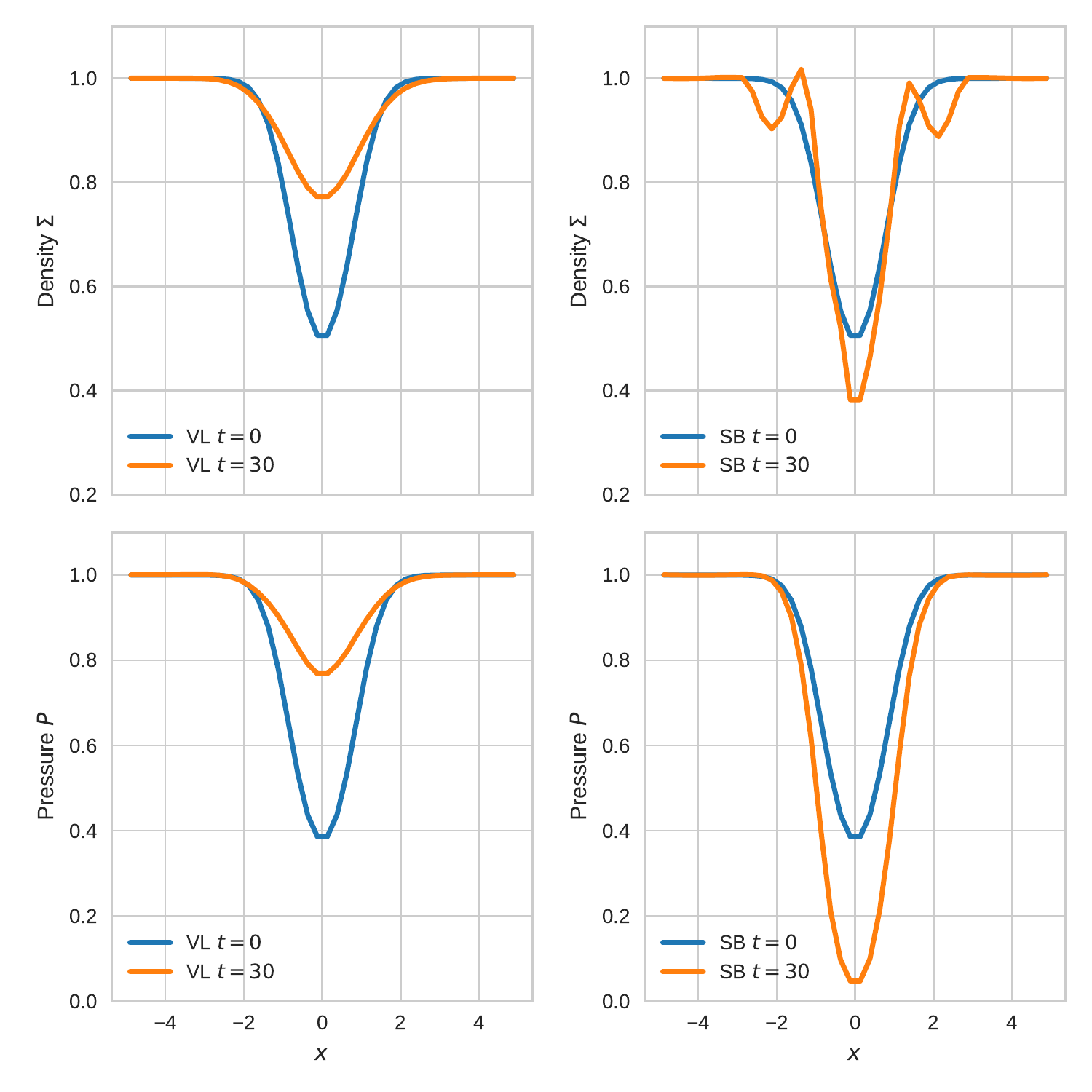}
    \caption{Cuts along the x-axis at $y=0.0$ for an isentropic vortex on a cartesian grid.
    \emph{Left (van-Leer):} density and pressure are smeared out.
    \emph{Right (Superbee):} density and pressure are both oversteepened with time.}
    \label{fig:isentropic_vortex}
\end{figure}

In Fig.~\ref{fig:isentropic_vortex} cuts of isentropic vortex tests are presented. The setup is
described in \citep{yee1999low,jung2018multi}. It has initial conditions
\begin{align}
  \varrho &= \varrho_{\infty} \left(1 - \frac{\left( \gamma-1 \right)}{8\gamma \pi^2} e^{1-r^2} \right)^{\frac{1}{\gamma - 1}} \\
  v_x &= \left( 1  - \frac{\chi}{2\pi} e^{\frac{1-r^2}{2}}\right) \\
  v_y &= \left( 1  + \frac{\chi}{2\pi} e^{\frac{1-r^2}{2}}\right) \\
  p &= p_{\infty} \left(\frac{\varrho}{\varrho_{\infty}}\right)^{\gamma},
\end{align}
with density $\varrho$, pressure $p$ and $v_x$, $v_y$ velocities in
the according directions. $r=\sqrt{x^2+y^2}$ is the distance from the center of the
vortex. The other setup parameters are summarized in Table~\ref{tab:vortex}.
\begin{table}[htbp]
  \caption{Setup for isentropic vortex}
  \centering
  \label{tab:vortex}
  \begin{tabular}{cc}
    \hline \hline
     setup variable & numerical value \\
    \hline
     background density $\varrho_{\infty}$  & $1.0$ \\
     background pressure $p_{\infty}$       & $1.0$ \\
     vortex strength $\chi$                 & $5.0$ \\
     adiabatic index $\gamma$               & $1.4$ \\
     domain size $x \times y$               & $[-5,5] \times [-5,5]$ \\
     resolution $N_x \times N_y$            & $40 \times 40$ \\
     limiters                               & van-Leer, Superbee \\
     simulation time $T$                    & $30$ \\
    \hline
  \end{tabular}
\end{table}
The results with van-Leer and Superbee limiting show a qualitative difference regarding
the evolution. With van-Leer the vortex is smeared out, while with Superbee the vortex is
strengthened.
\end{document}